\documentclass[journal]{IEEEtran}
\usepackage{amsmath,amsfonts}
\usepackage{algorithm}
\usepackage{algorithmic}
\usepackage{array}
\usepackage[caption=false,font=normalsize,labelfont=sf,textfont=sf]{subfig}
\usepackage{textcomp}
\usepackage{stfloats}
\usepackage{url}
\usepackage{verbatim}
\usepackage{graphicx}
\usepackage{cite}
\usepackage{xcolor}
\usepackage{lmodern}
\usepackage{multirow}
\usepackage{orcidlink}
\usepackage{booktabs}
\usepackage{fancyhdr}

\hypersetup{
hidelinks,
colorlinks=true,
linkcolor=black,
citecolor=blue,
urlcolor=black
}
\begin{document}

%  Too long. Maybe, GNN-enhanced Traffic Anomaly Detection for Next-Generation SDN-Enabled Consumer Electronics
\title{GNN-enhanced Traffic Anomaly Detection for Next-Generation SDN-Enabled Consumer Electronics}

\author{
Guan-Yan Yang$^{\orcidlink{0009-0002-2539-9057}}$,~\IEEEmembership{Graduate Member,~IEEE,} 
\and
Farn Wang$^{\orcidlink{0000-0002-0425-6500}}$,~\IEEEmembership{Member,~IEEE,} 
\and
Kuo-Hui Yeh$^{\orcidlink{0000-0003-0598-761X}}$,~\IEEEmembership{Senior Member,~IEEE}

\thanks{This work was partially supported by the Taiwan Academic Cybersecurity Center at the National Taiwan University of Science and Technology and by the National Science and Technology Council (NSTC) under Grants 114-2221-E-002-217, 114-2622-E-A49-022, 114-2221-E-A49-210, 114-2634-F-011-002-MBK, and 114-2923-E-194-001-MY3. Additional financial support was provided by National Taiwan University (NTU) and the NTU Core Consortium Project as part of the Higher Education Sprout Project by the Ministry of Education in Taiwan, under Grants NTU-CC-114L895501 and NTU-G0647.

Additional partial financial support was provided by the Department of Industrial Technology, Ministry of Economic Affairs, under the "2025 ITRI Advanced Research Program" (Grant No.: 114-EC-17-A-21-0337). Further partial support was also granted by the Hon Hai Research Institute in Taipei, Taiwan (Project No.: 114UA90042), and the Industry-Academia Innovation School, NYCU, Taiwan (Project No.: 113UC2N006).
The authors would like to express their gratitude for the financial support. 
\textit{(Corresponding author: Kuo-Hui Yeh \& Farn Wang.)}}
\thanks{Guan-Yan Yang and Farn Wang are with the Department of Electrical Engineering at National Taiwan University, Taipei 106319, Taiwan R.O.C. 
(e-mail: f11921091@ntu.edu.tw; farn@ntu.edu.tw).
}
\thanks{Kuo-Hui Yeh is with the Institute of Artificial Intelligence Innovation, National Yang Ming Chiao Tung University, No. 1001, Da Hsueh Road, East District, Hsinchu
City, 300093, Taiwan R.O.C., 
and also with the Department of Information Management, National Dong Hwa University, No. 1, Sec. 2, Da Hsueh Road, Shoufeng, Hualien, 974301, Taiwan R.O.C.
(e-mail: khyeh@nycu.edu.tw).}

% \thanks{Manuscript received ; revised .}
}

% \markboth{,~Vol.~9, No.~3, August~2024}%


\maketitle

\thispagestyle{fancy}
\fancyhf{}
  \renewcommand{\headrulewidth}{0pt} % 隱藏頁首分隔線
  \renewcommand{\footrulewidth}{0pt} % 隱藏頁腳分隔線
  
  % --- 設定頁首 (Header) ---
  % 我們在內容的開頭加上 \vspace*{-0.3in} 將整個區塊往上移動
  % 你可以自行微調 -0.3in 這個數值來達到想要的位置
  \fancyhead[L]{%
    \parbox{\textwidth}{\centering 
      \vspace*{-0.1in} 
      \fontsize{6}{9}\selectfont
      \textsf{ 
      Preprint. This article has been accepted for publication in IEEE Transactions on Consumer Electronics. This is the author's version which has not been fully edited and content may change prior to final publication.
      }
    }
  }

\fancyfoot[C]{%
    \parbox{\textwidth}{\centering \tiny
    \vspace*{0.2in} 
      \fontsize{6}{9}\selectfont
      \textsf{ 
      © 2025 IEEE. All rights reserved, including rights for text and data mining and training of artificial intelligence and similar technologies. Personal use is permitted,
but republication/redistribution requires IEEE permission. See https://www.ieee.org/publications/rights/index.html for more information.
      }
    }
  }
\renewcommand{\headrulewidth}{0pt}

\begin{abstract}
    Consumer electronics (CE) connected to the Internet of Things are susceptible to various attacks, including DDoS and web-based threats, which can compromise their functionality and facilitate remote hijacking. These vulnerabilities allow attackers to exploit CE for broader system attacks while enabling the propagation of malicious code across the CE network, resulting in device failures. Existing deep learning-based traffic anomaly detection systems exhibit high accuracy in traditional network environments but are often overly complex and reliant on static infrastructure, necessitating manual configuration and management. To address these limitations, we propose a scalable network model that integrates Software-defined Networking (SDN) and Compute First Networking (CFN) for next-generation CE networks. In this network model, we propose a Graph Neural Networks-based Network Anomaly Detection framework (GNN-NAD) that integrates SDN-based CE networks and enables the CFN architecture. GNN-NAD uniquely fuses a static, vulnerability-aware attack graph with dynamic traffic features, providing a holistic view of network security. The core of the framework is a GNN model (GSAGE) for graph representation learning, followed by a Random Forest (RF) classifier. This design (GSAGE+RF) demonstrates superior performance compared to existing feature selection methods. Experimental evaluations on CE environment reveal that GNN-NAD achieves superior metrics in accuracy, recall, precision, and F1 score, even with small sample sizes, exceeding the performance of current network anomaly detection methods. This work advances the security and efficiency of next-generation intelligent CE networks.
\end{abstract}


\section{Introduction}

\IEEEPARstart{T}{he} rapid expansion of the Internet of Things (IoT) has seamlessly integrated consumer electronics (CE) devices—such as smartphones, smartwatches, and laptops—into our daily lives, enabling remote access and connectivity across diverse sectors like e-healthcare, smart cities, and intelligent transportation \cite{wu2022state}. The CE market is projected to reach 2.873 billion users by 2025, driven by the capacity of nearly every device to generate and share data \cite{Statista2024ConsumerElectronics, 9199898}.

CE networks, composed of heterogeneous devices from various manufacturers, present unique challenges due to large-scale deployment, high device diversity, and limited computational resources \cite{wu2022state, 10.1109/TCE.2024.3367330}. Unlike traditional IT networks, CE devices such as smart home appliances and wearables require lightweight, secure, and low-latency communication \cite{10258346}. Their traffic is often encrypted, intermittent, and follows irregular patterns, complicating the task of network anomaly detection (NAD) \cite{adil2022ai}. Security breaches in CE can have severe consequences, including privacy invasion, financial loss, and physical safety risks, and compromised devices can be conscripted into botnets for large-scale attacks like DDoS campaigns \cite{9369044, Yang_2021, 10445503}.

While existing machine learning (ML) and deep learning (DL) methods for NAD have shown promise, they often suffer from time-consuming feature extraction processes and require extensive manual configuration, making them ill-suited for the dynamic nature of CE networks \cite{s21144884}. To overcome these limitations, advanced architectures like Compute First Networking (CFN) and Software-Defined Networking (SDN) are gaining traction. CFN integrates cloud, edge, and end-user computing resources to optimize performance \cite{li2019framework, tang9354741}, while SDN centralizes network control, providing a global view essential for robust security management \cite{chica2020security}. The synergy between SDN and GNN-NAD is particularly powerful; once an anomaly is detected, the SDN controller can be programmed to automatically install new flow rules to isolate a malicious device, thus creating a closed-loop system that moves seamlessly from detection to response.

This paper introduces a novel GNN-based Network Anomaly Detection (GNN-NAD) framework designed for SDN-enabled CE networks. The core novelty of our work lies in its holistic approach. Most NAD systems analyze only traffic statistics (the "how" of an attack), while traditional vulnerability analysis focuses only on static configurations (the "what"). Our framework uniquely combines both: we construct a static, vulnerability-aware attack graph that models potential exploit paths (the "what") and enrich it with real-time, dynamic traffic data (the "how"). This synthesis of static posture and dynamic behavior allows for a far more comprehensive and accurate security assessment than either approach could achieve alone.

The key contributions of this work are as follows:

\begin{itemize}

\item We propose a novel GNN-NAD framework that integrates a static attack graph with dynamic traffic features. Our specialized GNN model, GSAGE, is tailored to learn rich representations from this combined graph structure, outperforming standard GNN models.

\item We benchmark GNN-NAD against state-of-the-art (SOTA) NAD approaches, demonstrating its superior accuracy, precision, recall, and F1-score.

\item Through robustness analysis at various data sampling rates, we validate that our framework is highly effective for early detection, maintaining high accuracy even with limited data. Representing our work is probably widely used for real-world CE and IoT scenarios.

\end{itemize}

The rest of this paper is organized as follows: Section \ref{related_work} reviews related work. Section \ref{network_model} details the network model. Section \ref{approach} introduces the GNN-NAD framework. Section \ref{experiment} presents the experimental setup, evaluation metrics, and detection performance. Finally, Section \ref{conclusion} concludes the study and outlines directions for future research.

\section{Related Works} \label{related_work}
This section reviews prior work in CFN security and network anomaly detection, creating a foundation for our proposed framework by identifying existing research gaps.

\subsection{Security in CFN and Intelligent IoT}
Current research in Compute First Networking (CFN) primarily addresses integration architectures, resource scheduling, and security \cite{tang9354741, gong10419572, wang9293089}. Resource scheduling has been explored through centralized and decentralized management \cite{Yang2022}, microservice-based models \cite{yu2025microservice}, prioritization strategies \cite{Yang2024}, and cloud-edge frameworks \cite{zhou2023cloud}. Recent work in the broader field of AIoT has also addressed resource allocation challenges; for instance, Li et al. \cite{li2025} proposed an AI-driven task scheduling mechanism to manage computational resources in dynamic vehicular networks for smart warehousing, demonstrating the growing trend of integrating intelligence into resource management.

Security in CFN has been advanced through mechanisms like the CFN Watchdog for fault detection \cite{LIANG2021107873}, federated learning for secure operations \cite{zhao2020operation}, and blockchain for data integrity \cite{xu2019become}. However, as noted in \cite{10495806}, dedicated research on NAD specifically for CFN environments is still lacking. This represents a critical gap, as the distributed and dynamic nature of CFN introduces unique security challenges that traditional methods may not adequately address.

\subsection{Traditional and ML-based NAD}
Traditional NAD methods often rely on statistical features extracted from network flows. Vinayakumar et al. \cite{Vinayakumar8681044} employed Deep Neural Networks (DNNs) on the CIC-IDS2017 dataset, showing strong performance. Ning et al. \cite{ning9632825} improved cross-domain generalization through the integration of semi-supervised learning, transfer learning, and domain adaptation, validated on the USTC-TFC2016 dataset. Anbalagan et al. \cite{Anbalagan10144487} developed an intelligent intrusion detection system for 5G vehicular networks, employing enhanced convolutional neural networks and hyperparameter optimization on simulated traffic. Huang et al. \cite{HUANG2024103790} proposed a two-stage multi-label network attack detection method validated on the UNSW-NB15 dataset, demonstrating enhanced accuracy over existing techniques. Nie et al. \cite{nie9079682} examined abnormal traffic fluctuations in the Internet of Vehicles (IoV) by analyzing roadside units (RSUs), validated through self-simulated traffic anomalies. Other notable works in CE network include SEMI-GRU \cite{grualmahadin10288593}, which used a semi-supervised GRU for vehicular networks, and the work by Javeed et al. \cite{javeed2023intelligent}, which developed a CUDA-optimized BLSTM (Cu-LSTM) for SDN-based cyberattack detection. 

While effective, these approaches primarily analyze traffic statistics in isolation, ignoring the underlying network topology and device vulnerabilities. They can struggle to detect sophisticated attacks that mimic normal traffic patterns. Furthermore, many require extensive, well-labeled datasets for training, which are not always available in real-world CE environments.

\subsection{GNN-based and Advanced NAD}
To overcome the limitations of traditional methods, researchers have turned to Graph Neural Networks (GNNs), which can naturally model the relational structure of networks. 
Deng et al. \cite{Deng9919790} proposed a traffic graph with interval constraints and a spatial attention mechanism for label-limited intrusion detection, validated on three public datasets. Chang et al. \cite{chang2024embedding} addressed class imbalance by enhancing GraphSAGE and GAT algorithms with residual connections, showing effectiveness on UNSW-NB15 and CIC-IDS2017. 
\cite{GNNNIDSpujol2022unveiling} developed a GNN-based method for NAD that constructs host connection graphs using traffic flow features, though it faces high computational costs due to reliance on aggregated observations and traditional statistical features. This approach was validated on the CIC-IDS2017 dataset. 
Golchin et al. \cite{golchin2024sscl} achieved 96.54\% accuracy on the CIC-IDS2017 dataset, while Li et al. \cite{lis24072122} attained 99.9\% accuracy and a 95\% weighted recall rate by integrating contrastive learning with CNN and GRU models. Despite these promising results, a well-documented limitation of contrastive learning is its high computational overhead, which stems from the need for a large number of negative samples during training \cite{chen2020simple}.
Huang et al. \cite{HUANG2024103790} proposed a two-stage multi-label network attack detection method validated on the UNSW-NB15 dataset, demonstrating enhanced accuracy over existing techniques. Zeng et al. \cite{zeng2023towards} presented a "Human-in-the-Loop" framework that incorporates human expertise for intrusion detection, achieving 99.16\% accuracy with 28,100 manually labeled samples on the CIC-IDS2017 dataset.
Tran and Park \cite{tranapp14166932} developed a hybrid model integrating GCN and SAGEConv for graph-based feature learning, surpassing previous methods in accuracy and other performance metrics on the CIC-IDS2017 dataset. 
Recent related research by Islam et al. \cite{islam2025} introduces an explainable AutoML-driven scheme for intrusion prevention in Zero-Touch Networks, highlighting the trend towards automated and transparent security solutions. In 2025, Fu et al. \cite{fu2025fir} presented a GNN that uses flow interaction relationships for intrusion detection on consumer electronics within smart home networks. Concurrently, Pei et al. \cite{pei2025edge} proposed an edge intelligence-enabled network intrusion detection system tailored for Compute First Networking, emphasizing the move towards decentralized and resource-aware security architectures. Later, we introduced GSL-IDS \cite{Yang2025resilience}, a graph structure learning-based method for enhancing network resilience with network intrusion detection, which provides strong resilience against model adversarial attacks.

\subsection{Research Gap Summary} 
The existing literature reveals a clear gap: a lack of NAD frameworks that holistically integrate static vulnerability information with dynamic traffic analysis in a computationally efficient manner suitable for CE networks. Our GNN-NAD framework directly addresses this gap by creating a unified graph representation that captures both the "what" (vulnerabilities) and the "how" (traffic patterns) of network security.

\section{Our Network Model} \label{network_model}
The CFN is gaining recognition as an effective strategy for developing converged networks. Its architecture is defined by the separation of four key components: the computing resource pool (infrastructure), the routing plane, the control plane (orchestration and management layer), and the service plane (application layer). This separation ensures both simplicity and flexibility. In contrast, traditional networks are limited because each router can only perceive the status of its local network, lacking a comprehensive view of the entire network. This limitation creates significant challenges in developing robust defense mechanisms against network threats.

Integrating CFN with SDN gives the network a global perspective and centralized control capabilities. This combination facilitates easier access to network statistics. Within the CFN framework, the service plane offers computational services to users, enabling consumer electronics (CE) applications like smart homes, intelligent transportation systems, and mobile computing. The control plane handles routing, data transmission, and traffic monitoring through advanced application technologies. The routing plane consists of numerous CE switches and routers, while the computing resource pool includes various CE devices, such as smart devices, sensors, and other wireless technologies.

In our model, CE devices (such as smart TVs, wearables, home gateways, IoT sensors) are directly represented as nodes in the network, each with unique resource constraints and communication patterns. The SDN/CFN architecture enables dynamic, centralized management of these heterogeneous devices, supporting the specific needs of CE networks such as seamless device onboarding, real-time anomaly detection, and adaptive resource allocation. Our anomaly detection framework is designed to operate efficiently in such environments, considering the limited typical capabilities of CE devices.

\begin{figure}[!t]
\centering
\includegraphics[width=0.9\linewidth]{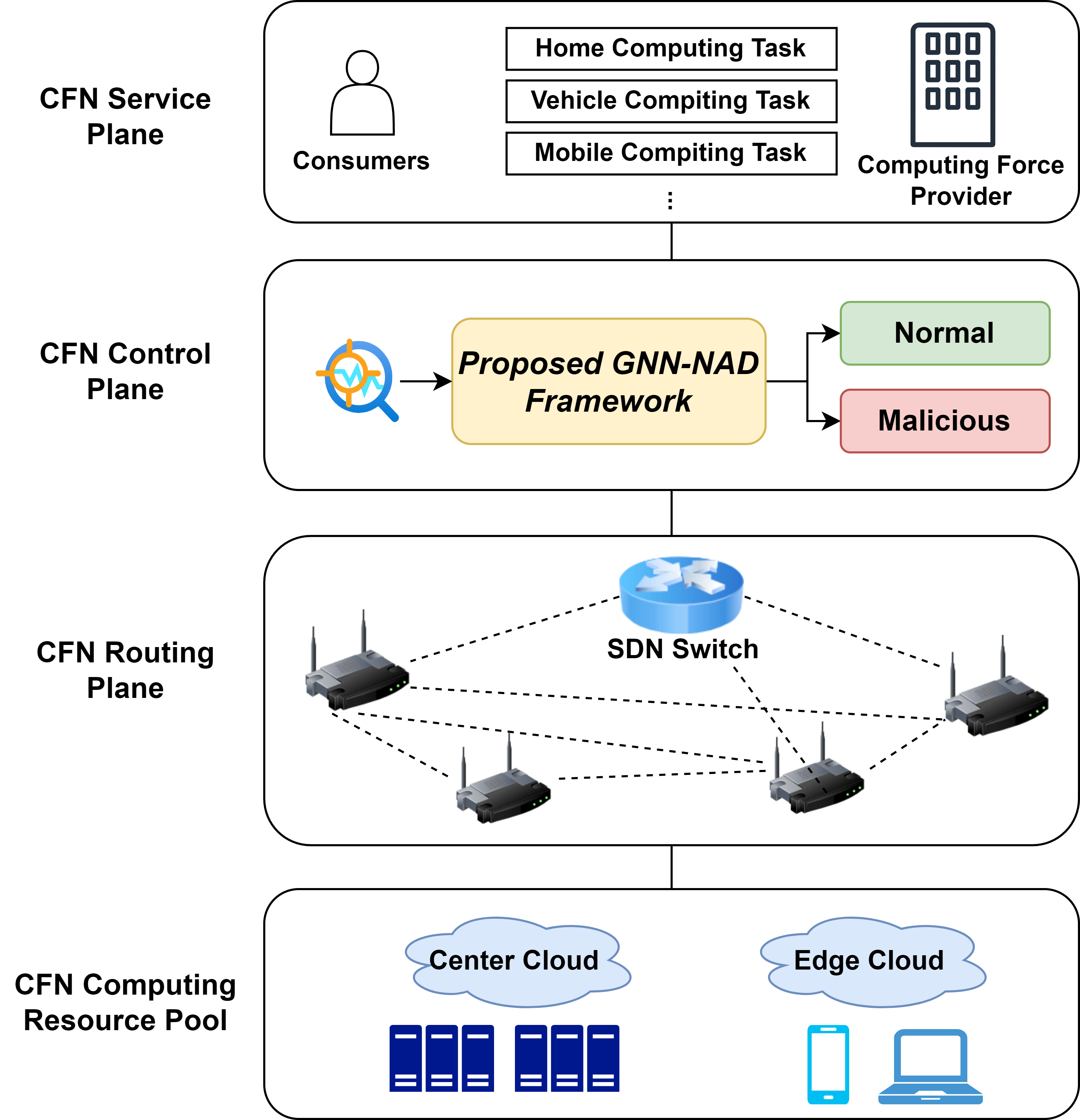}
\caption{Our Network Model}
\label{fig1}
\end{figure}

As illustrated in Figure \ref{fig1}, our model positions the proposed GNN-NAD framework within the CFN Control Plane. This placement is strategic: the control plane, managed by an SDN controller, has access to all network statistics and can dynamically manage network elements. This allows GNN-NAD not only to detect threats but also to orchestrate an immediate, automated response. For example, upon detecting a malicious device, the framework can instruct the SDN controller to install flow rules that isolate the device, effectively closing the loop from detection to mitigation.

Integrating CE, CFN, and SDN offers a streamlined solution for monitoring network traffic, enhancing the detection of attacks and suspicious activities. CFN and SDN provide a cohesive view of all devices and network elements, significantly improving the ability to monitor traffic and identify potential threats, attacks, and adverse events. Thus, CFN and SDN represent a promising direction for advancing CE networks.

\section{Proposed GNN-NAD Framework} \label{approach}
\begin{figure}[!t]
\centering
\includegraphics[width=0.95\linewidth]{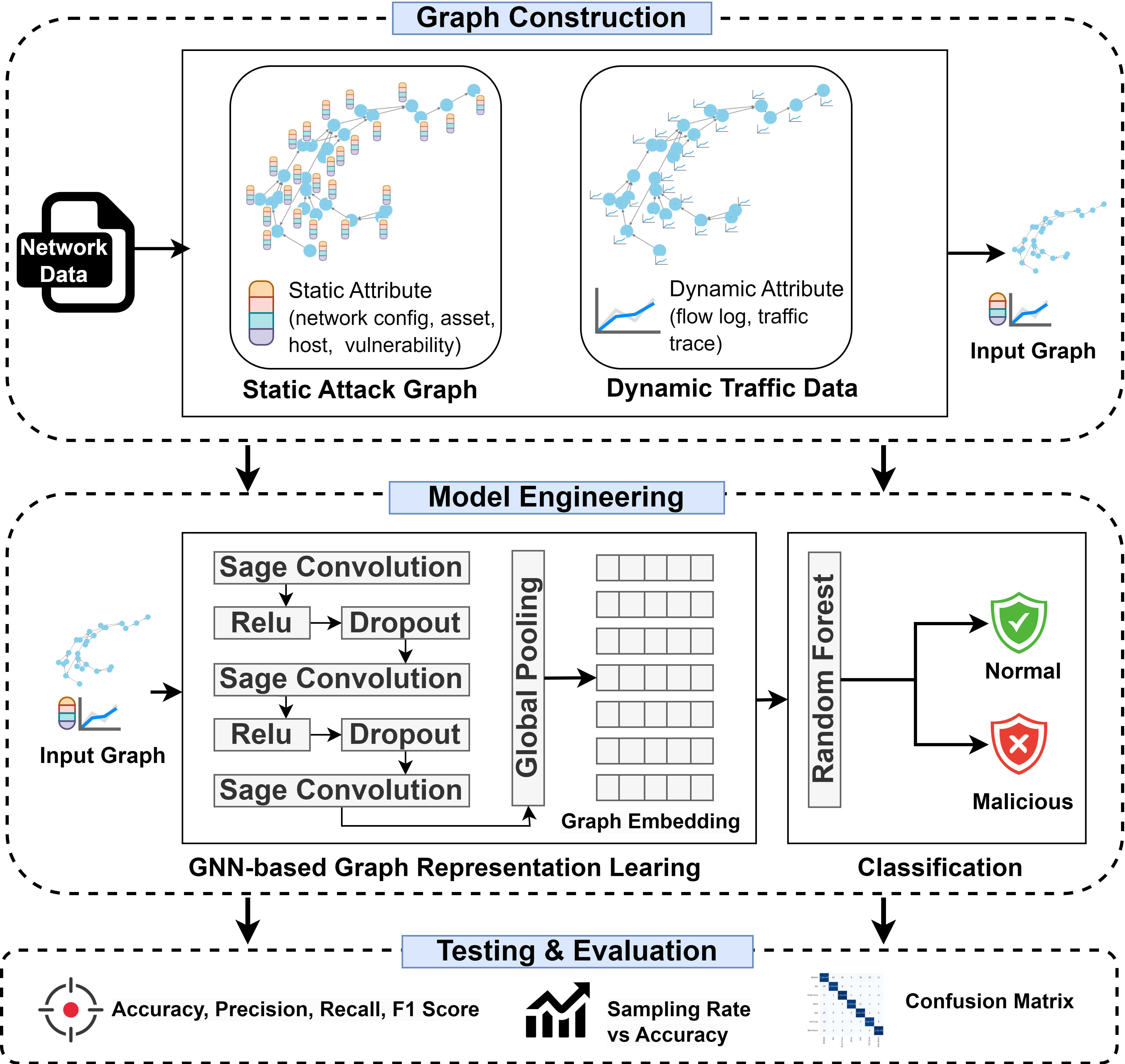}
\caption{Our Proposed Detection Framework}
\label{fig:approach}
\end{figure}

This section proposes a network anomaly detection method based on graph neural network embeddings (GNN-NAD), as illustrated in Figure \ref{fig:approach}. We first introduce the key components of GNN-NAD, including graph construction, GNN-based graph representation learning, and classification.

\subsection{Graph Construction}
\subsubsection{Static Attack Graph}

An attack graph (AG) is a directed acyclic graph (DAG) that represents the potential attack paths an adversary might exploit within a network by leveraging vulnerabilities \cite{konsta2024survey}. Based on network security analysis, several attack graph models have been proposed \cite{wang2019attacking, durkota2015optimal, presekal2023attack}. The Multi-host, Multi-stage Vulnerability Analysis Language (MulVAL) approach \cite{ou2005mulval} stands out for its scalability and low computational complexity. As a result, we employ MulVAL for AG generation and then design the AG encoding method to get a static attack graph (SAG) in our proposed GNN-based Network Attack Detection (GNN-NAD) framework.

We encode this logical AG into a Static Attack Graph (SAG) for our GNN. The SAG is represented as $\mathcal{SAG} = (V, E, F)$, where $V$ is the set of nodes, $E$ is the set of edges, and $F \in \mathbb{R}^{n \times D}$ is the node feature matrix. The features are derived from the text statements associated with each MulVAL node (such as `execCode(webServer,\_)'). As detailed in Algorithm \ref{alg1}, we use a bag-of-words approach to create a one-hot encoded vector $\mathbf{f}_v$ for each node $v$, where the vocabulary $\mathcal{C}$ is built from all unique tokens (words, predicates, parameters) across all node statements.

While more complex node embedding techniques like Word2Vec were considered, we chose the bag-of-words method for its simplicity, direct interpretability, and potential computational efficiency. This is a critical design choice for ensuring the framework remains lightweight and suitable for resource-constrained CE environments. Our experiments confirm that this straightforward encoding is highly effective for this task.

\begin{algorithm}[t]
\caption{Encode Attack Graph}
\label{alg1}
\begin{algorithmic}[1]
    \REQUIRE nodes $V$, edges $E$, statements $\{S_v \mid \forall v \in V\}$
    \ENSURE Encoded attack graph $SAG = (V, E, F)$

    \STATE Initialize $C_{\text{set}}$
    \FOR{$v \in V$}
        \FOR{$w \in S_v$}
            \STATE Add $w$ to $C_{\text{set}}$
        \ENDFOR
    \ENDFOR

    \STATE Convert $C_{\text{set}}$ to a list $C$
    \STATE Create an index map $\textit{map} \gets \{C[i] \to i\}$
    \STATE Let $D \gets |C|$
    \STATE Initialize a feature matrix $F[n][D]$ with zeros

    \FOR{$v \in V$}
        \FOR{$w \in S_v$}
            \STATE $i \gets \textit{map}[w]$
            \STATE $F[v][i] \gets 1$
        \ENDFOR
    \ENDFOR
\end{algorithmic}
\end{algorithm}

Since our focus is on generating a SAG, we collect the following types of static information:

\begin{itemize}
    \item \textbf{Network Configuration}: Includes the IP ranges, subnets, IP addresses, subnet masks, and gateway addresses of the covered network segments.
    \item \textbf{Network Assets}: Includes the hostnames, IP addresses, interface names, target network IP addresses, and subnet masks, as well as source and destination IPs and ports along with their associated transport protocols.
    \item \textbf{Network Hosts}: Captures the interface names, IP addresses, and their corresponding relationships with LAN/WAN connections.
    \item \textbf{Network Vulnerabilities}: Identifies potential CVEs collected from the National Vulnerability Database (NVD).
\end{itemize}

\subsubsection{Dynamic Traffic Data}

Dynamic measurements are typically presented in time series format, which is crucial for the real-time dynamic analysis of complex CE networks. In GNN-NAD, we have multiple actions, each representing either an attack or a benign action.
We represent dynamic measurements at time $t$ as $\mathbf{RDM}^t$, the real-time dynamic measurements of a network are expressed as: 
\[
\mathbf{RDM}^t = \{x^t_{\text{action}}, \forall \text{action} \in \text{actions}\}. 
\]
Here, \(x^t_{\text{action}} \in \mathbb{R}^K\)  and \(K\) is the number of measured variables.

These measurements are aggregated over discrete time windows. Let $t$ be a continuous time index. We aggregate features over a time window $[t_1, t_2]$.  Specifically, the $k$-th measurement at time $t$ can be calculated using the following formula:
\[
x^t_{\text{action},k} = \text{AGG}(x^{t_1}_{\text{action},k}, \ldots, x^{t_2}_{\text{action},k}), \quad t_1 \leq t \leq t_2 .
\]

After a number of these aggregation rounds, indexed by $k$, we can set \(X_{dyn}[action] = x_{action}\), i.e., obtain the final dynamic feature vectors for each observed action (such as a network flow). For clarity, $t$ refers to the fine-grained timestamp of a raw data point, while $k$ refers to the discrete update cycle or round in which aggregated data is processed.

\subsection{Integrating Static and Dynamic Features}
This step is crucial for fusing the "what" (static vulnerabilities) with the "how" (dynamic traffic). Dynamic traffic features, which are flow-based (such as containing source/destination IPs and ports), are mapped onto the nodes of the static graph. A dynamic feature vector from a specific traffic flow is mapped to a node $i$ in the SAG if the statement $S_i$ of that node contains corresponding network identifiers.

For example, a dynamic flow log showing traffic between IP `172.30.211.20' and `172.30.211.24' would have its feature vector (such as packet count, byte count) mapped to the specific nodes in the SAG that represent these two hosts. This is achieved by searching for the IPs in the node statements. The initial feature vector for each node $i$ is then formed by concatenating its static feature vector $F[i]$ with its mapped dynamic feature vector $X_{dyn}[i]$:
\[
h_i^{(0)} \gets \mathrm{Concat}(F[i], X_{dyn}[i]), \quad \forall i\in V
\]
This integrated feature vector $h_i^{(0)}$ serves as the input for the GNN.

\subsection{GNN-based Representation Learning}

To learn a comprehensive representation of the entire graph, we use our custom GNN model, GSAGE. As demonstrated in Figure \ref{fig:approach}, the GNN-NAD framework defines GSAGE as a model composed of three SAGE Convolution (SC) layers, each followed by a ReLU activation and a Dropout layer for regularization. This learned representation is then passed through a global pooling layer to produce a single embedding for the entire graph.

The distinction between our GSAGE model and a baseline GraphSAGE lies in its specific, streamlined architecture. While GraphSAGE is a general framework, GSAGE is an implementation optimized for our specific task. Key differences include:
\begin{itemize}
    \item \textbf{Architecture}: GSAGE uses a 3-layer stack with 256 hidden units, a configuration found to be optimal for capturing complex patterns in our vulnerability-traffic graphs without overfitting.
    \item \textbf{Efficiency}: The architecture is intentionally lean to ensure fast processing on resource-aware CE hardware. It is designed to handle the sparse but high-dimensional feature vectors generated by our graph construction process effectively.
\end{itemize}

The propagation rule for each SC layer $l$ is:
\[
h^{(l+1)}_i = \sigma\left(W^{(l)} \cdot \text{AGGR}\left(\{h^{(l)}_i\} \cup \{h^{(l)}_j, \forall j \in \mathcal{N}(i)\}\right)\right)
\]
where the propagation of features occurs across layers, with each node \( i \) represented by its feature vector \( h^{(l)}_i \) at layer \( l \). The transformation of these features relies on the trainable weight matrix \( W^{(l)} \) specific to that layer. To effectively capture information from neighboring nodes, the aggregation function denoted as \(\text{AGGR}\) is employed, which can take the form of mean or sum, depending on the desired outcome. Subsequently, an activation function \( \sigma \), such as the ReLU function, is applied to introduce non-linearity, facilitating complex pattern recognition within the graph. Furthermore, the set of neighbors for any given node \( i \) is represented by \(\mathcal{N}(i)\), which plays a crucial role in the aggregation process.

For other layers, Dropout regularization is applied to node representations:

\[
h^{(l)} = \text{Dropout}(h^{(l)}; p)
\]
where \( p \) is the dropout probability.

In the global pooling layer, we combine node embeddings into a single graph embedding:

\[
h_{\text{graph}} = \frac{1}{n} \sum_{i=1}^{n} h^{(L)}_i
\]
where \( n \) is the number of nodes and \( L \) is the final layer.

Moreover, the graph representation learning objective function uses cross-entropy loss.
After learning, the system will run the next phase, i.e., classification.

\subsection{Classification}
For our classification task, we chose to use the random forest algorithm. Random forest is a highly effective classifier known for its exceptional accuracy and strong ability to generalize. By combining predictions from multiple decision trees, it reduces the risk of overfitting, making it particularly suitable for high-dimensional data. Additionally, random forest includes an inherent feature selection mechanism. It also performs well with imbalanced datasets by adjusting sample weights to improve classification performance. Its strong resistance to noise further minimizes the impact of outliers on the model. For these reasons, we have selected the random forest classifier for this study.

\section{Experiment} \label{experiment}
This section presents the experiments and evaluation of the performance of our proposed GSAGE model. We detail the experimental setup, evaluation metrics, and discuss the results. Specifically, we investigate the effect of the number of samples used to update the graph on our detection performance of GSAGE. We also highlight the advantages of GSAGE in network traffic feature extraction. Furthermore, we demonstrate the performance of GNN-NAD in comparison to the SOTA NAD approach.

\subsection{Dataset and Preparation}
The CIC-IDS-2017 dataset \cite{sharafaldin2018toward} is one of the most widely utilized NAD datasets in recent years. It includes attack and benign traffic data collected over five days under simulated network conditions. For data preprocessing, we first removed all rows containing missing or non-numeric values, as these could negatively impact the model's performance. Given the imbalance in sample distribution, certain categories, such as infiltration and Heartbleed attacks, contained fewer samples. To address this imbalance, we excluded these minority categories from our experiments. We randomly selected 1,000 traffic samples from each remaining category. To better reflect real-world scenarios where benign traffic predominates, we randomly selected 9,000 benign samples, which represent approximately 56\% of the total dataset. Furthermore, we applied min-max normalization to the data before updating the graphs.

\subsection{Environment}

The proposed model is implemented using Python 3.8.20 and trained with PyTorch. All experiments are run on Docker. The experimental environment has been specifically designed to reflect the proposed CFN architecture using low-power, accessible hardware. The entire SDN topology is constructed using three Raspberry Pi computers. This setup provides a cost-effective yet powerful platform for emulating a CE network.

The roles of the Raspberry Pi devices are distributed as follows:
\begin{itemize}
\item Raspberry Pi 5 with 16GB RAM (Controller \& NAD): Hosts the ONOS SDN controller and runs the GNN-NAD detection framework. It manages the network and performs the data analysis.
\item Raspberry Pi 5 with 16GB RAM (SDN Switch): Configured with Open vSwitch to act as the forwarding element in the routing plane. It connects the traffic-generating device to the network and communicates with the ONOS controller via the OpenFlow protocol.
\item Raspberry Pi 4 with 8GB RAM (CE Device): Represents the CE devices in the "CFN Service Plane," such as smart TVs, wearables, or IoT sensors. It serves as the traffic source, generating diverse traffic patterns that simulate real-world CE device usage. As a node in the CFN Computing Resource Pool, its traffic patterns and resource constraints are designed to be representative of typical CE devices. It is used to replay the attack and benign scenarios from the CIC-IDS-2017 dataset.
\end{itemize}

\begin{figure}[!t]
\centering
\includegraphics[width=0.5\linewidth]{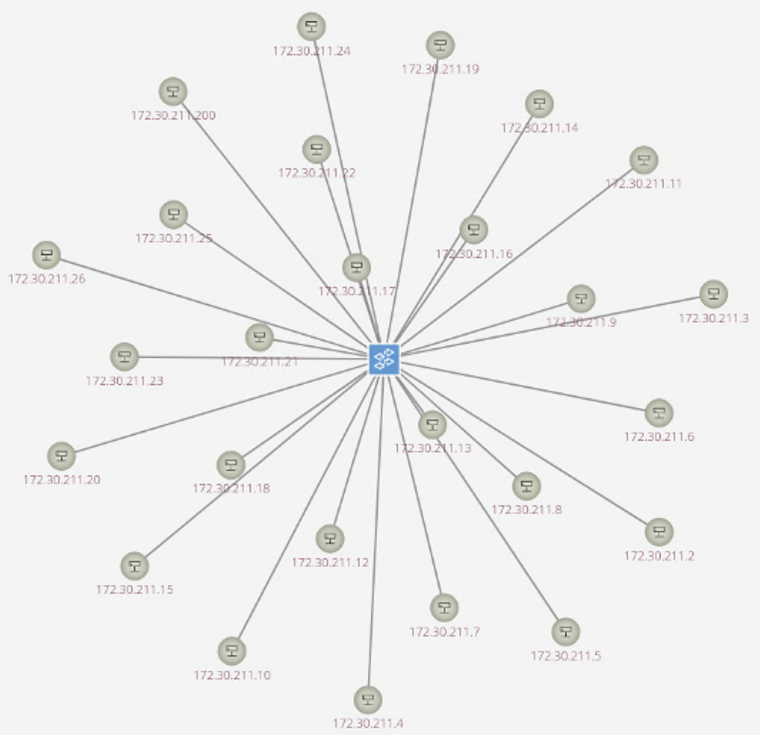}
\caption{Topology for our experiment. (Screenshot from Onos Controller.)}
\label{fig3}
\end{figure}

This physical topology, with the controller's view captured in Figure \ref{fig3}, allows us to evaluate the effectiveness and efficiency of our anomaly detection framework under conditions that closely resemble real-world, resource-aware CE deployments. The traffic patterns and attack scenarios are designed to reflect typical CE network usage, including bursty traffic and the introduction of malicious data flows, providing a robust testbed for our system.

\subsection{Metrics}
We evaluate GNN-NAD based on the following metrics:
\begin{align*}
    \text{Accuracy} &= \frac{TP + TN}{TP + TN + FP + FN}, \\
    \text{Recall} &= \frac{TP}{TP + FN}, \\
    \text{Precision} &= \frac{TP}{TP + FP}, \\
    F1\text{-score} &= 2 \times \frac{\text{Recall} \times \text{Precision}}{\text{Recall} + \text{Precision}},
\end{align*}
where TP, FP, TN, and FN represent true positives, false positives, true negatives, and false negatives, respectively. 

\subsection{Model Parameters}
%  random or not?
% [GY] Random. I missed adding.
We randomly partition the dataset by allocating 80\% of the samples to the training set, while reserving the remaining 20\% for the test set. Each experiment is repeated ten times to obtain average results. The parameters for the GSAGE, DNN, and GNN models are presented in Table \ref{tab:settings}. For the parameters of SOTA methods, we follow the settings in their article.

\begin{table}[h]
    \centering
    \caption{Parameter of Our Models}
    \label{tab:settings}
    \resizebox{\linewidth}{!}{
    \begin{tabular}{lcccc}
        \hline
        \textbf{Parameter} & \textbf{GSAGE} & \textbf{GCN} & \textbf{GraphSAGE} & \textbf{NN} \\
        \hline
        Optimizer & Adam & Adam & Adam & Adam \\
        Epoch & 100 & 100 & 100 & 100 \\
        Batch size & 32 & 150 & 150 & 256 \\
        Activation & Relu & Relu & Relu & - \\
        Dropout & 0.2 & 0.2 & 0.2 & 0.2 \\
        Hidden units & 256 & 64 & 64 & 256, 128, 64, 32 \\
        Num\_layers & 3 & 2 & 2 & 4 \\
        Graph\_pooling & sum & sum & sum & - \\
        Neighbor\_pooling & sum & - & mean & - \\
        \hline
    \end{tabular}
    }
\end{table}

% You should write some explanation for your experiment design and methods.
%[GY] Below:
\subsection{Experimental Methods}
For our experiment, we first examined the effectiveness of GSAGE+RF in traffic feature extraction and classification by contrasting it with a non-graph statistical flow feature method. We selected representative models from deep learning as comparison benchmarks. Additionally, we compared GSAGE+RF against two prominent GNN models to establish its superiority in classification contexts.
% We also implemented GNN+RF to assess GNN's GRL capabilities compared to GSAGE.

Furthermore, we evaluated GSAGE+RF against advanced NAD methods, including DNN, SEMI-GRU, Cu-LSTM, GNN-NIDS, FN-GNN, GSL-IDS, FIR-GNN, and NAAE-GCN. The first three utilize traditional approaches based on traffic statistical features, while the latter five incorporate structural features and statistics. We also investigated the influence of sampling rate on detection performance by systematically reducing the sample set of CICIDS2017, selecting 20\% of samples from each category.

\subsection{Results}

\subsubsection{Compare with Representative Models}

We first compared our graph-based approach (GSAGE+RF) against several baselines. This includes models using traditional statistical features extracted by CICFlowMeter (denoted with "CIC") and models using our graph construction method.

As shown in Table \ref{tab:bperformance}, our proposed GSAGE+RF model significantly outperforms all other configurations, achieving an accuracy of 99.69\%. Notably, when using statistical features from CICFlowMeter, our GSAGE model architecture still performs better than other model architectures (NN, GCN, GraphSAGE). This indicates the robustness of the GSAGE architecture itself. However, the most significant performance lift comes from combining GSAGE with our novel graph construction method, demonstrating the superiority of learning from integrated static and dynamic graph features.

\begin{table}[!t]
    \centering
    \caption{Performance Comparison with Representative Methods}
    \label{tab:bperformance}
    \resizebox{\linewidth}{!}{
    \begin{tabular}{lccccc}
        \toprule
        \textbf{Methods} & \textbf{Accuracy} & \textbf{Recall} & \textbf{Precision} & \textbf{F1-score}\\
        \midrule
        NN(CIC)       & 0.9509 & 0.9536 & 0.9594 & 0.9565 \\
        GCN(CIC)      & 0.6622 & 0.6972 & 0.7007 & 0.6990 \\
        GraphSAGE(CIC) & 0.7991 & 0.8350 & 0.8129 & 0.8238 \\
        GSAGE(CIC)    & 0.8809 & 0.8917 & 0.8938 & 0.8939 \\
        NN            & 0.6388 & 0.6728 & 0.6811 & 0.6769 \\
        GCN           & 0.7534 & 0.7911 & 0.7452 & 0.7831 \\
        GraphSAGE     & 0.7541 & 0.8011 & 0.7707 & 0.7856 \\
        GSAGE         & 0.9888 & 0.9928 & 0.9873 & 0.9900 \\
        GSAGE+RF & \textbf{0.9969} & \textbf{0.9983} & \textbf{0.9961} & \textbf{0.9972} \\
        \bottomrule
    \end{tabular}
    }
\end{table}

\begin{figure}[!t]
\centering
\includegraphics[width=0.9\linewidth]{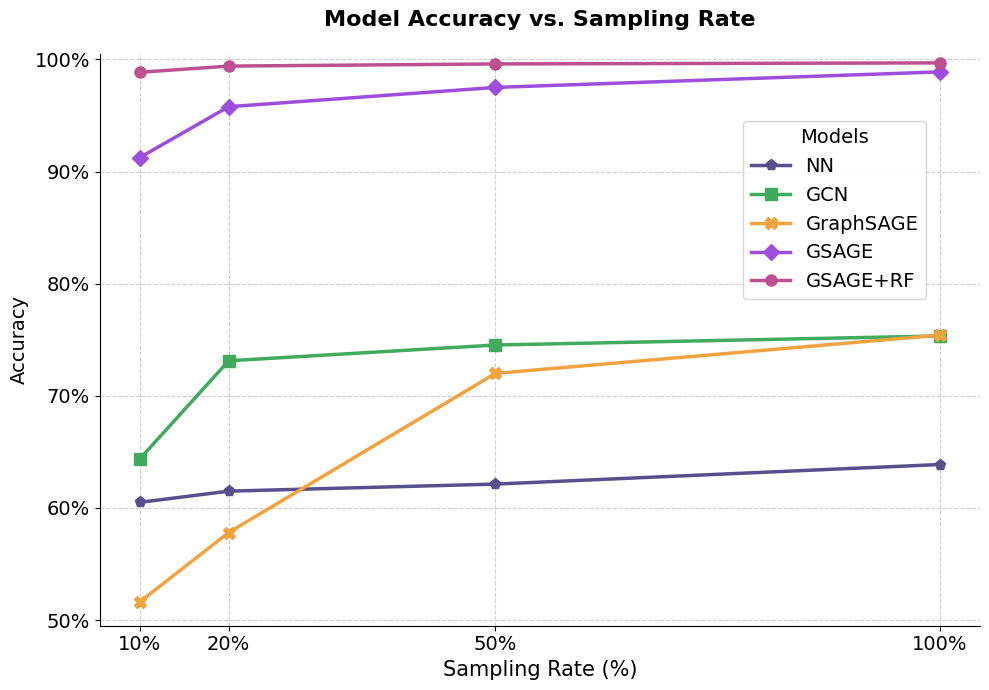}
\caption{Compare with widely used NN and GNN models with different sample rates with our graph classification method.}
\label{fig:baseline_sample}
\end{figure}

To further assess model performance under different data conditions, we trained models using various sampling rates, with accuracy results shown in Figure \ref{fig:baseline_sample}. GSAGE+RF consistently outperformed baseline methods across all sampling rates, achieving an accuracy of 98.85\% even at a 10\% sampling rate. 

\begin{figure}[!t]
\centering
\includegraphics[width=1.0\linewidth]{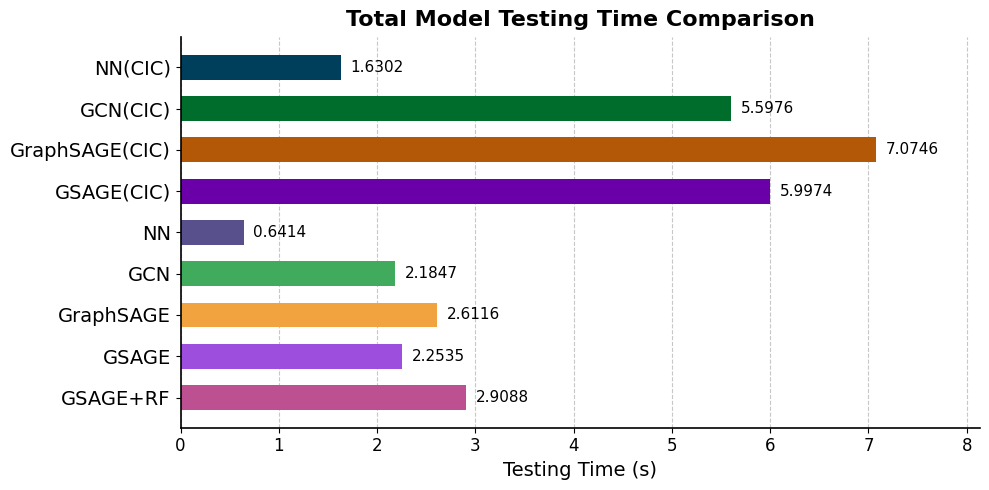}
\caption{Compare Testing Time with Baseline methods.}
\label{fig:baseline_time}
\end{figure}

We also analyzed model testing time, defined as the execution time on a pre-constructed graph or pre-processed features (Figure \ref{fig:baseline_time}). The results show two key trends. First, our graph-based methods are substantially more efficient than the CICFlowMeter-based approaches. For instance, inference with our GSAGE model took 2.25s, a marked improvement over the 7.07s for GraphSAGE with CICFlowMeter features. Second, among the different architectures, NN-based models were the fastest, and GSAGE was consistently quicker than the standard GraphSAGE model.
Moreover, the initial static graph construction is a one-time offline cost (taking 4.12 seconds in our experiments) and is not included in this recurring testing time, making our overall approach highly efficient for consumer electronic networks that require lightweight and high accuracy.

\subsubsection{Compare with SOTA NAD}

\begin{figure}[!t]
\centering
\includegraphics[width=0.9\linewidth]{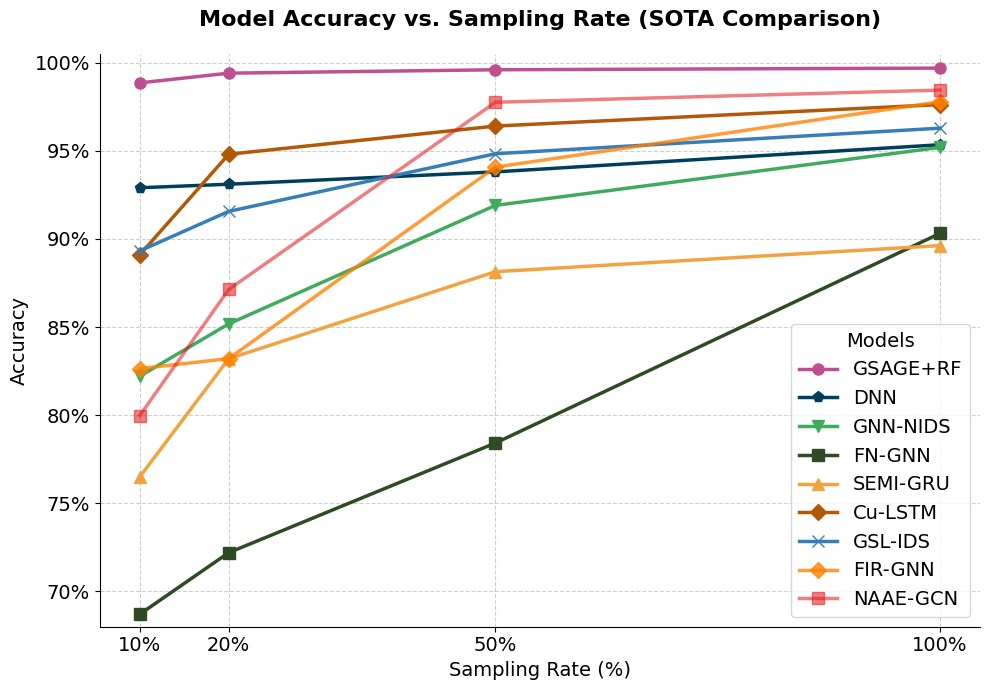}
\caption{Compare with SOTA NAD methods with different sample rates.}
\label{fig:sota}
\end{figure}

% Fewer sample size is your strength; you should mention that.
%[GY] Updated
We evaluated the performance of GSAGE+RF against various SOTA methods in traffic feature extraction and graph classification. As shown in Table \ref{tab:sota_performance}, GSAGE+RF outperformed all SOTA approaches across four key metrics. Additionally, we trained our model on datasets sampled at different rates, with accuracy results illustrated in Figure \ref{fig:sota}. When the sample size was limited (10\%), most SOTA methods struggled to achieve satisfactory performance, with only simple models like DNN surpassing 90\% accuracy. In contrast, GSAGE+RF achieved 98.85\% accuracy and consistently outperformed other methods across all sampling rates. These results demonstrate that our GNN-NAD approach, powered by GSAGE+RF, can effectively detect network traffic anomalies with minimal training data.

\begin{table}[!t]
    \centering
    \caption{Performance Comparison with SOTA Methods}
    \label{tab:sota_performance}
    \resizebox{\linewidth}{!}{
    \begin{tabular}{lcccc}
        \toprule
        \textbf{Methods} & \textbf{Accuracy} & \textbf{Recall} & \textbf{Precision} & \textbf{F1-score} \\
        \midrule
        DNN \cite{Vinayakumar8681044}      & 0.9534 & 0.9550 & 0.9619 & 0.9585 \\
        GNN-NIDS \cite{GNNNIDSpujol2022unveiling}       & 0.9519 & 0.9341 & 0.9839 & 0.9583 \\
        FN-GNN \cite{tranapp14166932}        & 0.9031 & 0.9080 & 0.9211 & 0.9145 \\
        SEMI-GRU \cite{grualmahadin10288593}    & 0.8962 & 0.9044 & 0.9105 & 0.9075 \\
        Cu-LSTM \cite{javeed2023intelligent} & 0.9762 & 0.9832 & 0.9744 & 0.9788 \\
        GSL-IDS \cite{Yang2025resilience} & 0.9628 & 0.9511 & 0.9822 & 0.9664 \\
        FIR-GNN \cite{fu2025fir} & 0.9778 & 0.9750 & 0.9854 & 0.9802 \\
        NAAE-GCN \cite{pei2025edge} & 0.9844 & 0.9850 & 0.9862 & 0.9871 \\
        GSAGE+RF  & \textbf{0.9969} & \textbf{0.9983} & \textbf{0.9961} & \textbf{0.9972} \\
        \bottomrule
    \end{tabular}
    }
\end{table}

\section{Conclusion and Future Work} \label{conclusion}
In this paper, we introduced GNN-NAD, a novel network anomaly detection framework tailored for next-generation SDN-enabled CE networks. Our approach formulates NAD as a graph classification problem by uniquely fusing a static attack graph, which captures network vulnerabilities (the "what"), with dynamic traffic data (the "how"). The core of our framework, a combination of a streamlined GSAGE model and a Random Forest classifier, proved highly effective. Experiments conducted on a realistic testbed of CE devices demonstrated that GNN-NAD achieves state-of-the-art performance, maintaining exceptional accuracy even with limited training data. This makes it a practical and powerful solution for securing heterogeneous and resource-constrained CE environments.

This work opens several exciting avenues for future research.
\begin{itemize}
\item \textbf{Proactive and Predictive Defense:} The static attack graph contains rich information about potential attack paths. This could be leveraged not just for detection but for proactive defense, such as identifying critical vulnerabilities to patch to disrupt the most likely attack chains.
\item \textbf{Federated and Online Learning:} To enhance privacy and adaptability, a federated version of GNN-NAD could be developed to train models across multiple CE networks without sharing raw data. Furthermore, transitioning to an online learning model would allow the system to adapt to new threats continuously without the need for periodic offline retraining.
\item \textbf{Explainable AI (XAI):} A major challenge with GNNs is their "black box" nature. Integrating XAI techniques, such as GNNExplainer, would provide invaluable insights into why the model flags certain activities as malicious, making the system more transparent and trustworthy for security analysts.
\end{itemize}

By pursuing these directions, we can further enhance the security, intelligence, and autonomy of future CE networks.

\section*{Acknowledgment}

Thanks to editors and anonymous reviewers for their valuable comments.
We appreciate the technical support from the Speech AI Research Center at the National Yang Ming Chiao Tung University. 
% Thanks, Zhao Min Chen (Security Researcher at CyCraft Technology), for lending us a Raspberry Pi 4.
Guan-Yan Yang is grateful to the National Science and Technology Council (NSTC) in Taiwan for the graduate research fellowship (NSTC-GRF) and to Professor Hung-Yi Lee for co-hosting his Ph.D research project. 

\bibliographystyle{IEEEtran}
\bibliography{IEEEabrv,ref}

% \newpage
 
\vspace{8pt}

% \end{IEEEbiography}


\begin{IEEEbiography}[{\includegraphics[width=1in,height=1.25in,clip,keepaspectratio]{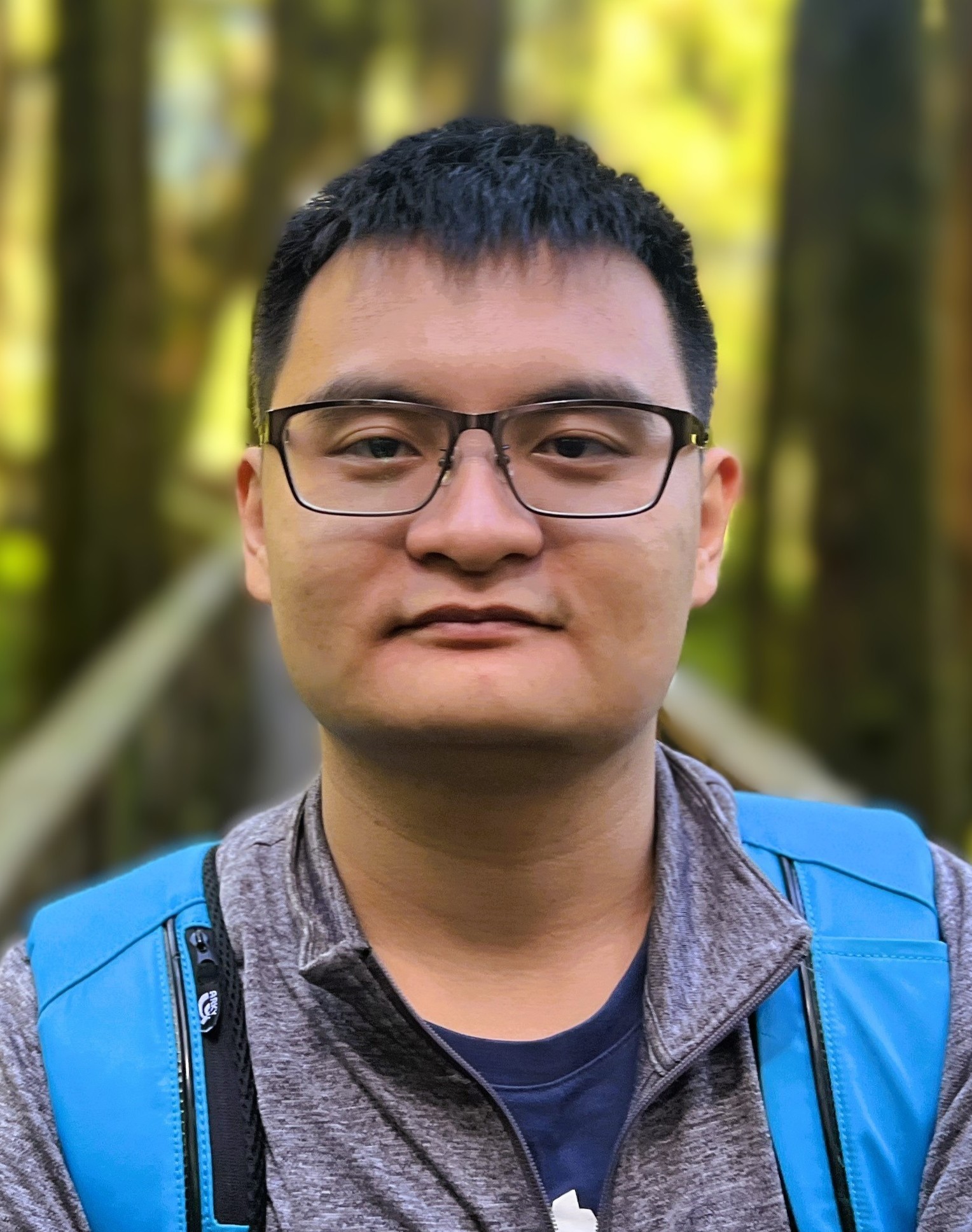}}]{Guan-Yan Yang} (Graduate Member, IEEE) 
    received a Bachelor's degree from the Department of Information Management at National Dong Hwa University, Hualien, Taiwan, in 2022. 
    He is currently pursuing a Ph.D. in the Department of Electrical Engineering at National Taiwan University, Taipei, Taiwan.
    In 2023, he worked as a Software Engineer at the Design Technology Platform in the Research and Development division of the Taiwan Semiconductor Manufacturing Company. Since 2024, he has been a researcher at the Taiwan Academic Cybersecurity Center and the Institute of Information Science at Academia Sinica in Taiwan. In 2024, he received a scholarship from the Norman and Lina Chang Foundation in the USA. That same year, he was awarded a graduate research fellowship in the information security category from the National Science and Technology Council. Additionally, he won the 7th and Taiwan Star Award (first place in Taiwan) in the world security competition HITCON CTF.
    % His research has been presented at USENIX Security and IEEE NETWORK.
    His research interests include security, safety, deep learning, generative AI, the Internet of Things, formal verification, and software testing.
    Mr. Yang is a member of the IEEE Computer Society, the IEEE Reliability Society, the IEEE Communication Society, the IEEE Consumer Technology Society, and SEAT.
\end{IEEEbiography}

% \begin{IEEEbiography}[{\includegraphics[width=1in,height=1.25in,clip,keepaspectratio]{figures/fig1.png}}]{Zhao Min Chen}
%     received the B.S. degree in Department of Computer Science and Information Engineering from National Taiwan Normal University in 2022, and the MS degree in Electrical Engineering from National Taiwan University in 2024. 
%     He worked as an Adjacent Security Researcher at CyCraft AI Lab, Taipei, Taiwan, from September 2022 to July 2024.
%     Starting in August 2024, he has been worked as a Security Research Fellow at CyCraft AI Lab, Taipei, Taiwan.
%     His research interests include cybersecurity, bluetooth, vulnerability detection, reverse engineering and fuzzing.
%     His recent research on fuzzing for Bluetooth Protocol was presented at USENIX Security 2024.
%     Mr. Chen is a member of Balsn CTF Team in Taiwan and has participated in several CTFs. 
%     Mr. Chen has won 7th place in HITCON CTF 2023, 3rd place in DEFCON 31, and 7th place in DEFCON 32 represent Taiwan. 
% \end{IEEEbiography}
\vspace{-33pt}
\begin{IEEEbiography}[{\includegraphics[width=1in,height=1.25in,clip,keepaspectratio]{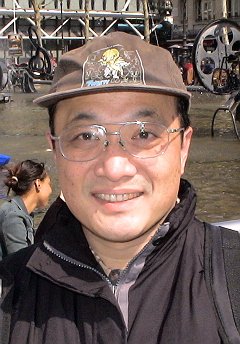}}]{Farn Wang} (Member, IEEE) 
    is a Full Professor at the Department of Electrical Engineering, National Taiwan University.
    He received the B.S. degree in Electrical Engineering from National Taiwan University in 1982 and the M.S. degree in Computer Engineering from National Chiao-Tung University in 1984. 
    He completed his Ph.D. in Computer Science at the University of Texas at Austin in 1993. 
    He is a founding member and chairman of the Steering Committee of the International Symposium on Automated Technology for Verification and Analysis (ATVA) from 2003 to 2022 and has served on the ATVA advisory committee since 2022.  
    He was also an Associate Editor of FMSD (International Journal on Formal Methods in System Design), Springer-Verlag.   
    His research interests include formal verification, model-checking, software testing, security, verification automation, AI, and language models.
    He has been named a World's Top 2\% Scientists in a career-long list by Stanford University since 2020.
\end{IEEEbiography}

\vspace{-33pt}
\begin{IEEEbiography}[{\includegraphics[width=1in,height=1.25in,clip,keepaspectratio]{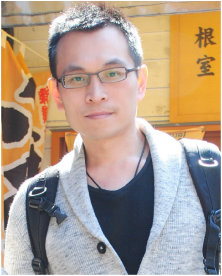}}]{Kuo-Hui Yeh} (Senior Member, IEEE) 
serves as a professor at the Institute of Artificial Intelligence Innovation, National Yang Ming Chiao Tung University, Hsinchu, Taiwan. 
Prior to this appointment, he was a professor in the Department of Information Management at National Dong Hwa University, Hualien, Taiwan, from February 2012 to January 2024. 
Dr. Yeh earned his M.S. and Ph.D. degrees in Information Management from the National Taiwan University of Science and Technology, Taipei, Taiwan, in 2005 and 2010, respectively. 
He has contributed over 150 articles to esteemed journals and conferences, covering a wide array of research interests such as IoT security, Blockchain, NFC/RFID security, authentication, digital signatures, data privacy and network security. 
Furthermore, Dr. Yeh plays a pivotal role in the academic community, serving as an Associate Editor (or Editorial Board Member) for several journals, including the Journal of Information Security and Applications (JISA), Human-centric Computing and Information Sciences (HCIS), Symmetry, Journal of Internet Technology (JIT) and CMC-Computers, Materials \& Continua. 
In the professional realm, Dr. Yeh is recognized as a Senior Member of IEEE and holds memberships with ISC2, ISA, ISACA, CAA, and CCISA. His professional qualifications include certifications like CISSP, CISM, Security+, ISO 27001/27701/42001 Lead Auditor, IEC 62443-2-1 Lead Auditor, and ISA/IEC 62443 Cybersecurity Expert, covering fundamentals, risk assessment, design, and maintenance specialties. 
\end{IEEEbiography}

\vfill

\end{document}